\documentclass[aip,jcp,reprint]{revtex4-1}
\usepackage{graphicx}% Include figure files
\usepackage{dcolumn}% Align table columns on decimal point
\usepackage{bm}% bold math
\usepackage[utf8]{inputenc}
\usepackage[T1]{fontenc}
\usepackage{mathptmx}
\usepackage{etoolbox}
\usepackage{multirow} % <-- Add this line to use \multirow command
\usepackage{xcolor}  % For colored text

\makeatletter
\def\@email#1#2{%
 \endgroup
 \patchcmd{\titleblock@produce}
  {\frontmatter@RRAPformat}
  {\frontmatter@RRAPformat{\produce@RRAP{*#1\href{mailto:#2}{#2}}}\frontmatter@RRAPformat}
  {}{}
}%
\makeatother
\begin{document}

\preprint{AIP/123-QED}

\title[Y. Song and X. Wu]{Pressure-Induced Structural and Dielectric Changes in Liquid Water at Room Temperature}
% Force line breaks with \\
\author{Yizhi Song}
\affiliation{Department of Physics, Temple University, Philadelphia, Pennsylvania 19122, USA}
\email{yizhi.song@temple.edu}

\author{Xifan Wu}
\affiliation{Department of Physics, Temple University, Philadelphia, Pennsylvania 19122, USA}
\affiliation{Institute for Computational Molecular Science, Temple University, Philadelphia, Pennsylvania 19122, USA.}

\date{\today}% It is always \today, today,
             %  but any date may be explicitly specified

\begin{abstract}
Understanding the pressure-dependent dielectric properties of water is crucial for a wide range of scientific and practical applications. In this study, we employ a deep neural network trained on density functional theory data to investigate the dielectric properties of liquid water at room temperature across a pressure range of 0.1 MPa to 1000 MPa. We observe a nonlinear increase in the static dielectric constant \(\varepsilon_{0}\) with increasing pressure, a trend that is qualitatively consistent with experimental observations. This increase in \(\varepsilon_{0}\) is primarily attributed to the increase in water density under compression, which enhances collective dipole fluctuations within the hydrogen-bonding network as well as the dielectric response. Despite the increase in \(\varepsilon_{0}\), our results reveal a decrease in the Kirkwood correlation factor \(G_K\) with increasing pressure. This decrease in \(G_K\) is attributed to pressure-induced structural distortions in the hydrogen-bonding network, which weaken dipolar correlations by disrupting the ideal tetrahedral arrangement of water molecules.
\end{abstract}

\maketitle

\section{\label{sec:intro}Introduction} 

Water's high static dielectric constant, resulting from the high molecular polarity of water molecules, makes it a universal solvent. Effectively screening electrostatic interactions between positive and negative charges, this property is crucial for the dissolution of ionic crystals, such as salts, and significantly influences water's solvent behavior in both natural and industrial applications.\cite{Fer97, Bakker08, Tobias08, Marcus09, Stir13, Ding13} In geological and biological contexts, water is often subjected to high pressures, such as those found in deep oceanic environments or within cellular structures under stress. These varying pressure conditions can induce changes in water's structure and dynamics, which in turn impact its static dielectric constant, thereby affecting the solubility of minerals and the nature of chemical reactions.\cite{Helge81,Wein05} Furthermore, according to the Debye-Hückel theory,\cite{Debye1923} the static dielectric constant and its pressure derivatives are related to the infinite-dilution limiting slopes of the thermodynamic properties of electrolyte solutions, including activity and osmotic coefficients. Therefore, a comprehensive understanding of the pressure-dependent dielectric constant of water is essential for gaining deeper insights into its behavior in both natural and engineered systems. 

Over the decades, extensive experimental research has been conducted to investigate the dependence of water's dielectric properties on temperature and pressure.\cite{Uematsu80, Donald90, Wasser95, Fern95} Within the temperature range between water's normal freezing and boiling points, a small positive slope in the dielectric constant as a function of pressure has been observed, and the partial derivative of the dielectric constant with respect to pressure at constant temperature decreases as pressure increases. In 1979, Bradley and Pitzer developed a model to describe the dielectric constant as a function of temperature and pressure,\cite{Bradley1979} with parameters adjusted to reproduce existing experimental data. Since then, several additional correlations have been proposed to describe the dielectric constant across various temperature and pressure ranges.\cite{Pitzer83, Donald90, Fer97} However, all of these equations are empirical, fitted to available experimental data, and may yield nonphysical results when extrapolated beyond the range of experimental data, which is limited to pressures below 0.5 GPa.\cite{Fern95} Despite these sustained efforts, a quantitative, molecular-level understanding of the increase in water's dielectric constant with pressure at room temperature remains challenging to achieve.

Recent advances in computer simulations, particularly molecular dynamics (MD) simulations, have provided valuable insights into the dielectric properties of water. However, accurately predicting dielectric constants remains challenging due to the sensitivity of the results to the specific potentials used. Classical MD simulations typically rely on empirical force fields, which are optimized to reproduce accurate values for pressure and energy within specific temperature and pressure ranges. For instance, the SPC/E water model\cite{Berendsen1987} has been shown to be capable of reproducing the experimental value of the dielectric constant over a density range of 0.326–0.998 g/cm$^3$, while also accurately predicting the temperature trend of the Kirkwood g-factor.\cite{Guillot93} Despite these successes, the reliability of classical force fields diminishes when applied beyond their parameterized conditions. Moreover, such models often assume a rigid water molecule, neglecting the flexibility of hydrogen bonds and fluctuations in electric polarizability, both of which are essential for a precise description of water's dielectric properties. To overcome these challenges, \textit{ab initio} molecular dynamics (AIMD), based on density functional theory (DFT),\cite{Car85,Hohen64,Kohn65} has become an essential tool for predicting water's properties from first principles. Unlike empirical force fields, AIMD constructs the potential energy surface on-the-fly from DFT calculations, without the need for parameterization. AIMD provides a more accurate representation of water's behavior under various conditions,\cite{Swartz13, Guidak17, Rozsa20} and has proven to be a powerful method for studying the static dielectric constant.\cite{Pan13, Hou20, Zhang23} 

While AIMD offers high accuracy in studying the dielectric properties of water, its application is significantly limited by the high computational cost. The long-range nature of dipole-dipole interactions, coupled with the disordered structure of liquid water, requires large-scale models with hundreds of molecules and simulation timescales extending beyond nanoseconds.\cite{Krish21,Zhang16} Such extensive simulations remain a challenge with current computational resources. Recent advancements in machine learning, particularly the deep potential molecular dynamics (DPMD) method,\cite{Zhang18} present a promising solution to these challenges. By training machine learning models on first principles data, these models can accurately learn potential energy surfaces, enabling MD simulations with the precision of first principles methods but the efficiency of empirical force fields. Additionally, the centers of electronic orbitals, described using maximally localized Wannier functions,\cite{Marzari97} can be predicted via the Deep Wannier neural network model.\cite{Zhang20} Machine learning techniques have been successfully employed to obtain the dielectric constant of supercritical water under a wide range of pressures and temperatures,\cite{Hou20} demonstrating their capability for high-pressure studies. However, the variation of the dielectric constant with pressure at ambient temperature remains largely unexplored with this state-of-the-art approach.

In this study, we utilize deep neural networks (DNNs) trained on data from DFT calculations based on the strongly constrained appropriately normed (SCAN) functional,\cite{Sun15,Sun16} to investigate the dielectric properties of liquid water across a pressure range of 0.1 MPa to 1000 MPa at room temperature. Our results show a nonlinear increase in the static dielectric constant \(\varepsilon_{0}\) with pressure, which is qualitatively consistent with experimental observations. This increase can be primarily attributed to the nonlinear rise in water density under compression. As the liquid compresses, a greater number of water molecules occupy each unit volume, thus enhancing collective dipole fluctuations within the hydrogen-bonding network as well as the dielectric response. Moreover, the strengthening of hydrogen bonds under pressure contributes to an increase in the dipole moment of individual water molecules, further increasing the dielectric constant. We also observe a modest increase in the electronic contribution in the high-frequency limit, \(\varepsilon_{\infty}\), which can be attributed to the enhanced strength of interband transitions right across the bandgap, though its contribution to the overall dielectric constant remains minimal. Despite the observed increase in \(\varepsilon_{0}\), our results reveal a decrease in the Kirkwood correlation factor \(G_K\) with pressure. This decrease is attributed to pressure-induced structural distortions in the hydrogen-bonding network, which disrupt the ideal tetrahedral arrangement of water molecules. While the directional hydrogen bonds become stronger, these distortions weaken angular correlations among water dipoles. Overall, our findings offer crucial insights into the behavior of liquid water under pressure, with potential implications for diverse fields, such as materials science and environmental studies.

The rest of this paper is organized as follows: In Sec. II, we describe the methods employed for the machine learning models and the molecular dynamics simulations used to investigate the dielectric properties of water. Our results are presented and discussed in Sec. III. Finally, our conclusions are summarized in Sec. IV.

\section{\label{sec:methods}Methods}

To simulate the dielectric constant of liquid water accurately, it is essential to incorporate long-range dipole-dipole interactions into the potential energy surface model. For this purpose, we employed the Deep Potential Long-Range (DPLR) method\cite{Zhang22}, which combines both short-range and long-range contributions to construct the potential energy surface. The short-range interactions are modeled as in the standard deep potential model,\cite{Zhang18} while the long-range electrostatic interactions are calculated from the electrostatic energy between spherical Gaussian charges located at ionic and electronic sites. Additionally, we trained a Deep Wannier DNN model\cite{Zhang20} to accurately partition the electronic charge density into contributions from the dipole moments of individual water molecules. This model predicts the centroids of the maximally localized Wannier centers (WCs) for each atom, allowing precise calculation of molecular dipole moments.\cite{Marzari12} Both the DPLR and Deep Wannier models were trained on data from DFT calculations using the SCAN functional, and were implemented using the DeePMD-kit package.\cite{Wang18} This dual DNN framework enables us to compute the static dielectric constant of water with DFT-level accuracy while maintaining computational efficiency. 

Training configurations, each containing 64 water molecules, were generated using an active learning procedure.\cite{Zhang20} The energies and forces of these configurations were computed using DFT within the Quantum ESPRESSO (QE) package.\cite{Giannozzi09} Electron-nuclei interactions were modeled with the pseudopotential method, specifically employing Hamann-Schluter-Chiang-Vanderbilt (HSCV) pseudopotentials.\cite{Hamann1979, Vanderbilt85} A plane-wave energy cutoff of 150 Ry was applied throughout the calculations. The electronic ground state was considered converged when the energy difference between consecutive self-consistent electronic iterations was less than $1 \times 10^{-6}$ Ry. After convergence, unitary transformations were performed to convert the occupied Bloch orbitals into maximally localized Wannier functions, using the Wannier90 code.\cite{Pizzi20} The WCs associated with each water molecule were then calculated from these maximally localized Wannier centers (MLWCs).

We first trained a Deep Wannier DNN model using the atomic configurations and their corresponding WCs as input. The architecture of the model consisted of embedding and fitting networks with sizes (25, 50, 100) and (100, 100, 100), respectively. Training was performed over $2 \times 10^{6}$ steps using the Adam stochastic gradient descent optimizer. The learning rate was set to decay exponentially, starting from an initial value of $1 \times 10^{-2}$ and reducing to $5.6 \times 10^{-8}$ by the end of the training. Following this, the DPLR model was trained using the same set of atomic configurations, combined with the corresponding energies, forces, and the Deep Wannier model. The size of the DPLR model's embedding and fitting networks was (25, 50, 100) and (240, 240, 240), respectively. The DPLR model was trained with the Adam stochastic gradient descent optimizer for $2 \times 10^{6}$ steps, with an initial learning rate of $1 \times 10^{-3}$, which decayed exponentially to $1 \times 10^{-8}$.

For the simulations, we performed eleven independent DPLR MD simulations of water, covering pressures from 0.1 to 1000 MPa, within periodic simulation cells containing 512 water molecules. The simulations were performed using the Large-scale Atomic/Molecular Massively Parallel Simulator (LAMMPS),\cite{Plimpton95} integrated with the DeePMD-kit package.\cite{Wang18} Long-range electrostatic interactions were handled using the particle-particle-particle-mesh (PPPM) method.\cite{Hockney89} All simulations were carried out in the isobaric-isothermal (NPT) ensemble at a temperature of 330 K, with metallic electric boundary condition applied. The chosen simulation temperature of 330 K, slightly higher than the experimental temperature of 298 K, compensates for the SCAN functional’s known overestimation of the melting temperature of ice.\cite{Piaggi21} Each simulation was run for approximately 20 nanoseconds (ns) to ensure the convergence of the dielectric constant values, with the first 100 picoseconds (ps) discarded to allow for equilibration.

To evaluate the accuracy of the DPLR simulations, we compared the forces, energies, and molecular dipoles of water predicted by the DNN model with those obtained from DFT calculations, using configurations that were not included in the DNN model’s training set. The DNN model successfully reproduced the DFT results with high accuracy. Specifically, the root-mean-squared errors (RMSE) for the energy and atomic forces, relative to DFT, were 0.44 meV/atom and 0.06 eV/Å, respectively, consistent with the typical accuracy achieved in deep potential training. (See Supplemental Material for more details). Additionally, the RMSE for the water molecular dipole was 0.06 Debye, which is significantly smaller than the average water molecular dipole of approximately 3 Debye, further validating the precision of the DNN model in predicting molecular dipole moments.

\section{\label{sec:results}Results and Discussion}

\subsection{\label{sec:dielectric}Pressure-Induced Variations in the Static Dielectric Constant}

It is widely accepted that the high static dielectric constant of water is closely associated with the correlated fluctuations of molecular dipoles within its hydrogen-bonding network. By analyzing the MD trajectories generated using our DNN model, we calculated the static dielectric constant \(\varepsilon_{0}\) of water under various pressure conditions. In MD simulations with periodic boundary conditions, the dielectric constant of an isotropic and homogeneous fluid can be determined from the fluctuations in the total dipole moment \(\mathbf{M}\), as described by linear response theory:\cite{Neumann84,Sharma07}

\begin{equation}
\varepsilon_{0} = \frac{4\pi}{3k_{B}T V}\left( \left\langle \mathbf{M}^{2} \right\rangle - \langle \mathbf{M} \rangle^{2} \right) + \varepsilon_{\infty},
\label{eq:fluctuation}
\end{equation}
where \(k_{B}\) is the Boltzmann constant, \(T\) is the temperature, \(V\) is the volume of the simulation box, and \(\varepsilon_{\infty}\) is the electronic contribution. \(\mathbf{M}\) was computed as the sum of the dipole moments of individual water molecules, \(\mathbf{\mu}_i\), where \(\mathbf{\mu}_i = 6 \mathbf{R}_O + \mathbf{R}_{H 1} + \mathbf{R}_{H 2} - 2 \sum_{j=1}^4 \mathbf{R}_{Wj}\), expressed in atomic units assuming e = 1. Here, \(\mathbf{R}_O\), \(\mathbf{R}_{H 1}\), and \(\mathbf{R}_{H 2}\) are the coordinates of the oxygen and hydrogen atoms of molecule \(i\), and \(\mathbf{R}_{Wj}\) represents the \( j \)-th center of the maximally localized Wannier functions associated with molecule \(i\).

To calculate the electronic contribution \(\varepsilon_{\infty}\), we employed density functional perturbation theory (DFPT).\cite{Baroni01} Unlike the static dielectric constant \(\varepsilon_0\), which is highly sensitive to molecular dipole fluctuations, \(\varepsilon_{\infty}\) exhibits significantly smaller fluctuations,\cite{Pan14} enabling its convergence with only a few tens of MD configurations. As indicated in Table 1, \(\varepsilon_{\infty}\) increases from 1.88 to 2.11 over the pressure range studied. This increase can be attributed to the enhanced strength of interband transitions right across the bandgap.\cite{Pan14} It has been reported that although the bandgap itself increases with pressure, the delocalization of the valence band edge leads to greater overlap with the conduction band minimum, thereby enhancing the interband transition strength.\cite{Pan14} While the contribution of \(\varepsilon_{\infty}\) to the total dielectric constant \(\varepsilon_0\) is relatively small, it remains a noticeable effect.

\begin{figure}
\includegraphics[width=3.4in,height=2.51in]{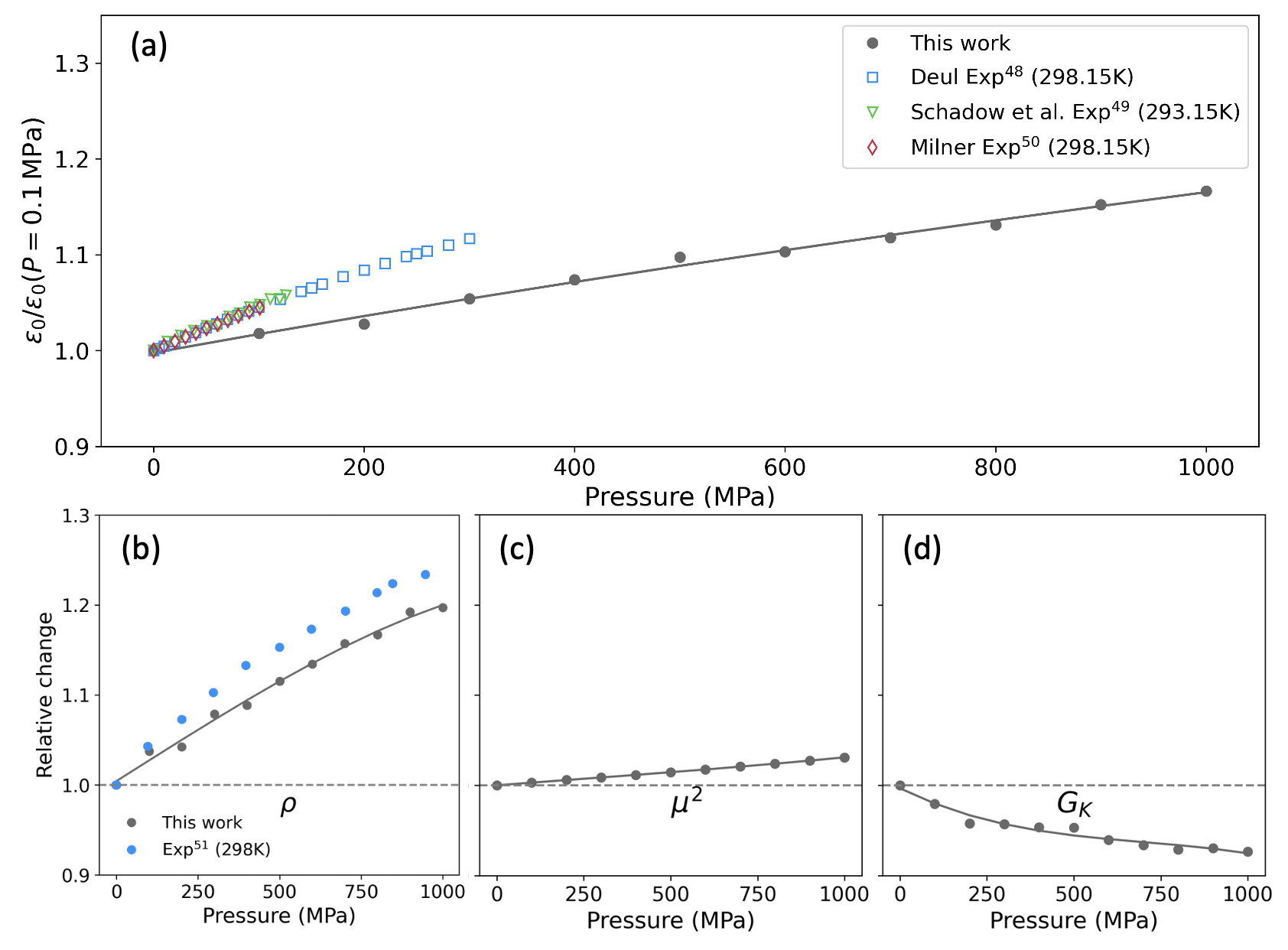}
\caption{\label{fig:dielectric}Pressure-dependent dielectric properties of water. (a) Relative static dielectric constant from this work and experiments.\cite{Deul84, Schadow69, Milner55} (b-d) The relative changes of water density \(\rho\) from this work and experiments,\cite{Koster69} water molecular dipole moment squared \(\mu^2\), and Kirkwood correlation factor \(G_K\), as a function of pressure. All results are normalized by their respective values at 0.1 MPa.}
\end{figure}

\begin{table*}
\caption{\label{tab:dielectric}Variation of the static dielectric constant \(\varepsilon_{0}\), density \(\rho\), dipole moment \(\mu\), Kirkwood factor \(G_{K}\), and the electronic contribution \(\varepsilon_{\infty}\) of water as a function of pressure.}
\begin{ruledtabular}
\begin{tabular}{lccccccccccc}
\multirow{2}{*}{Parameters} & \multicolumn{11}{c}{Pressure (MPa)} \\
& 0.1 & 100 & 200 & 300 & 400 & 500 & 600 & 700 & 800 & 900 & 1000 \\
\hline
\(\varepsilon_{0}\) & 99.82 & 101.57 & 102.54 & 105.17 & 107.17 & 109.49 & 110.05 & 111.50 & 112.84 & 114.96 & 116.35 \\
\(\rho\) (g/mL) & 1.062 & 1.102 & 1.107 & 1.145 & 1.156 & 1.184 & 1.204 & 1.229 & 1.239 & 1.266 & 1.271 \\
\(\mu\) (Debye) & 2.968 & 2.972 & 2.976 & 2.980 & 2.985 & 2.989 & 2.994 & 2.998 & 3.003 & 3.008 & 3.013 \\
\(G_{K}\) & 3.406 & 3.336 & 3.262 & 3.259 & 3.247 & 3.247 & 3.200 & 3.180 & 3.164 & 3.169 & 3.156 \\
\(\varepsilon_{\infty}\) & 1.88 & 1.91 & 1.95 & 1.96 & 1.98 & 2.01 & 2.03 & 2.05 & 2.06 & 2.09 & 2.11 \\
\end{tabular}
\end{ruledtabular}
\end{table*}

Our results, presented in Fig.~\ref{fig:dielectric}(a) and Table 1, reveal an increase in \(\varepsilon_{0}\) with increasing pressure, following a nonlinear trend where the rate of increase gradually diminishes at higher pressures. This behavior is qualitatively consistent with experimental observations. However, it is important to acknowledge the inherent limitations of the SCAN functional, which is known to overestimate dielectric constants. For instance, our predicted dielectric constant of water at 0.1 MPa is 99.82, which exceeds the experimental value of 78.36. This overestimation is consistent with previous studies that also employed the SCAN functional,\cite{Krish21, Zhang23} and can be attributed to the self-interaction error inherent in the SCAN functional, which tends to overstrengthen hydrogen bonding. This effect results in a slightly overstructured liquid water, which has been widely reported.\cite{Chen17, Xu20, Piaggi21} 

To gain a deeper understanding of the mechanisms driving the observed increase in the dielectric constant, we employ the Kirkwood formalism,\cite{Kirkwood1939} a powerful framework for understanding the dielectric properties of polar liquids, explicitly accounting for molecular dipole correlations. Within this formalism, the static dielectric constant \(\varepsilon_{0}\) is expressed as: 

\begin{equation}
\varepsilon_{0} = \frac{\rho \mu^{2} G_{K}}{3 k_{B} T \varepsilon_{vac}} + \varepsilon_{\infty},
\label{eq:Kirkwood}
\end{equation}
where \( \rho \) is the number density of the system, \( \mu \) is the average dipole moment per water molecule, \( G_{K} \) is the Kirkwood correlation factor, which quantifies the overall angular correlations among water dipoles, and \( \varepsilon_{vac} \) is the vacuum permittivity. This formulation allows us to decompose the contributions to the dielectric constant into three distinct factors: the number density \( \rho \), the intrinsic molecular properties (dipole moment \( \mu \)), and the molecular dipolar correlations (the Kirkwood correlation factor \( G_{K} \)).

We performed a detailed analysis of water's behavior at room temperature under varying pressures. As shown in Fig.~\ref{fig:dielectric}(b-d) and summarized in Table 1, we systematically examined the contributions of various parameters to the pressure-dependent changes in \(\varepsilon_{0}\), beyond the electronic contributions. The primary factor driving the increase in \(\varepsilon_{0}\) is the nonlinear rise in number density \( \rho \), which is consistent with the trend reported in prior research.\cite{Koster69, Hayward67} As pressure increases, the compression of the liquid leads to a higher number of water molecules within the same volume. This increase in density enhances collective dipole fluctuations, \(\left\langle \mathbf{M}^{2} \right\rangle \), within the hydrogen-bonding network, as well as the dielectric response, as described by Eqs. (1) and (2). Meanwhile, the shortening of O-O distances under pressure strengthens hydrogen bonds, contributing to a slight increase in the molecular dipole moment \( \mu \), further increasing \(\varepsilon_{0}\). (See Supplemental Material for the distribution of \( \mu \) under different pressures). However, despite the overall increase in \(\varepsilon_{0}\), the Kirkwood correlation factor \( G_{K} \) decreases with pressure, indicating that angular correlations among water dipoles become weaker. Therefore, the pressure-induced changes in the static dielectric constant \(\varepsilon_{0}\) is governed by a competition between the increasing density, which drives the increase in dielectric constant, and the diminishing dipolar correlations, which counteract this rise. The structural distortions and their impact on dipolar correlations will be discussed in more detail in the following subsections.

\subsection{\label{sec:structure}Structural Distortions in the Hydrogen-Bonding Network Under Pressure}

\begin{figure}
\includegraphics[width=3.5in,height=3.54in]{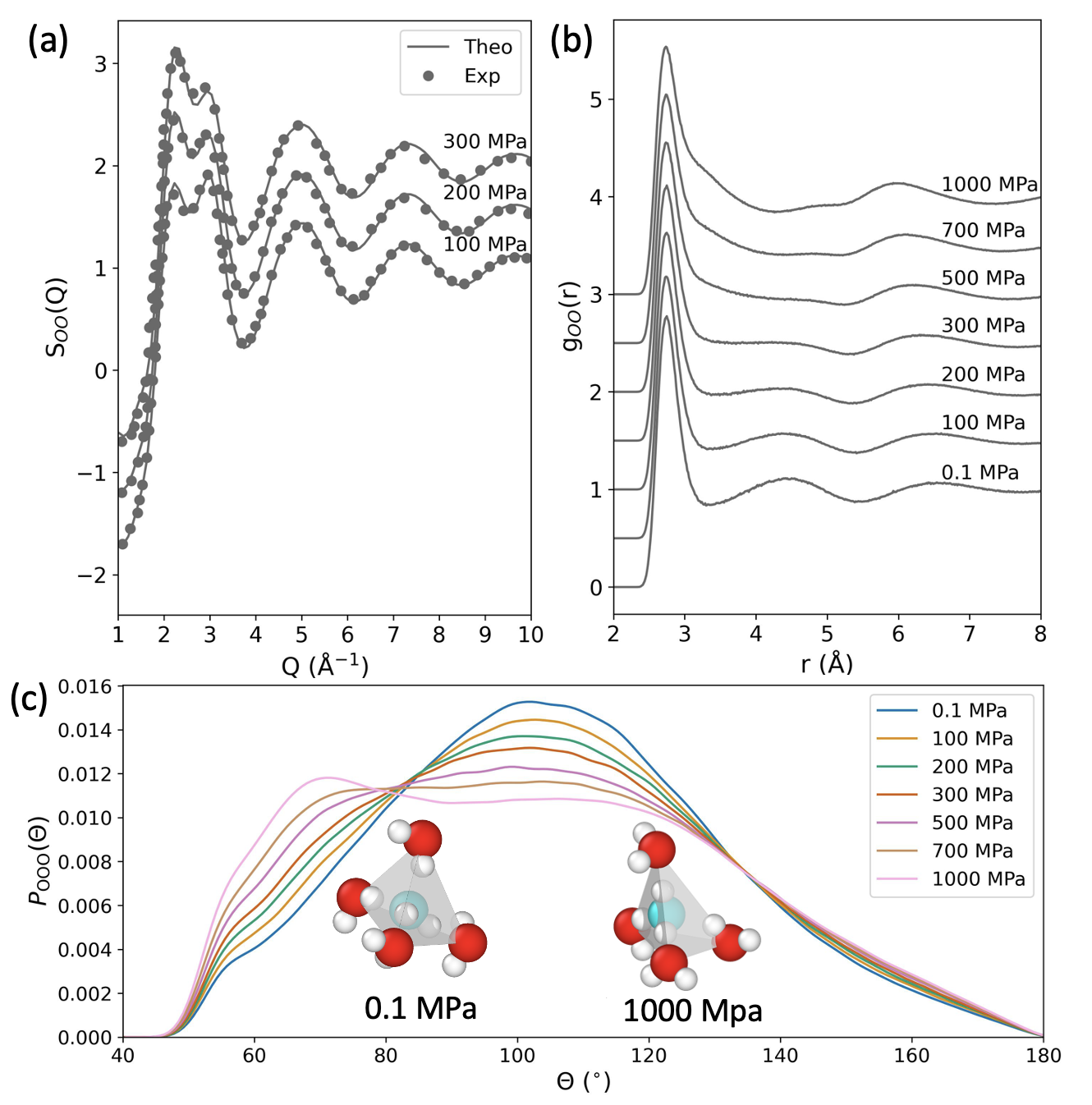}
\caption{\label{fig:structure}
Real-space and reciprocal-space structural variations of water under different pressures. (a) Experimental\cite{Skinner16} and theoretical structure factors \(S_{OO}(Q)\) for oxygen atoms in water at various pressures. All curves are shifted vertically for visual clarity. (b) Calculated O-O radial distribution function, \(g_{OO}(r)\), showing the effect of pressure on the spatial arrangement of water molecules. (c) O-O-O triplet angle distribution \(P_{\text{OOO}}(\theta)\) for water molecules under different pressures. The insets show the tetrahedral structure of water molecules at 0.1 MPa and 1000 MPa, respectively, indicating the pressure-induced distortions in the hydrogen-bonding network. Several pressure conditions are omitted for visual clarity.}
\end{figure}

As pressure increases, the intermolecular distances in liquid water decrease, leading to a more compact molecular arrangement. Experimental techniques such as X-ray diffraction (XRD) and neutron diffraction are commonly used to accurately analyze the molecular structure of liquid water. In particular, the structure factor \(S_{OO}(Q)\) provides valuable information about spatial correlations among water molecules in reciprocal space. This factor is derived from the interference patterns measured in diffraction experiments, describing how incident radiation is scattered by the molecular arrangement. Figure~\ref{fig:structure}(a) presents both experimental and theoretical \(S_{OO}(Q)\) under various pressure conditions. The theoretical structure factors were computed by Fourier transforming the real-space correlation functions into reciprocal space. The excellent agreement between the experimental and theoretical results confirms the accuracy of our simulations in capturing the structural features of water under pressure. Previous studies have shown that applied pressure significantly alters the structure factor of water.\cite{Skinner16, Okhulkov94} As shown in Figure~\ref{fig:structure}(a), increasing pressure results in a more pronounced first peak of \(S_{OO}(Q)\), while the second peak becomes attenuated. These changes suggest significant structural rearrangements in the liquid water, indicating substantial distortions in the tetrahedral arrangement of water molecules at high pressures.

To gain further insight into the structural changes occurring in real space, we analyze the oxygen-oxygen (O-O) radial distribution function (RDF), \(g_{OO}(r)\), across the same range of pressures, as shown in Fig.~\ref{fig:structure}(b). The RDF provides a quantitative measure of the spatial correlations between oxygen atoms in water molecules. As pressure increases, we observe a notable inward shift of the second and third coordination shells, along with an increased population of interstitial water molecules between the first and second shells. These rearrangements contribute directly to the observed increase in water density under pressure. Notably, the position of the first peak in \(g_{\text{OO}}(r)\) remains nearly unchanged, indicating that the primary hydrogen bonds are rather robust. However, the shift and attenuation of the second and later peaks indicate significant distortions in the tetrahedral structure of water.

Under ambient conditions, the electronic orbitals of water molecules adopt sp\textsuperscript{3} hybridization, resulting in a near-ideal tetrahedral structure. A water molecule at the center of a tetrahedron typically donates two H-bonds and accepts two from neighboring water molecules at the four vertices of the tetrahedron. This structure is characterized by an O-O distance of approximately 4.5 Å, as reflected in the second peak of the radial distribution function \(g_{OO}(r)\), shown in Fig.~\ref{fig:structure}(b). The tetrahedral geometry is further indicated by the O-O-O triplet angle distribution, \(P_{\text{OOO}}(\theta)\), which exhibits a primary peak near 104.5\textdegree{}, close to the ideal tetrahedral angle of 109.5\textdegree{}, as shown in Fig.~\ref{fig:structure}(c). With increasing pressure, however, we observe a significant decrease in the intensity of the second peak in \(g_{OO}(r)\), indicating substantial distortions in the tetrahedral network. This structural rearrangement is further confirmed by changes in \(P_{\text{OOO}}(\theta)\), where the primary peak weakens with increasing pressure. Additionally, a new peak emerges at smaller angles around 70\textdegree{}, becoming dominant at higher pressures, as shown in Fig.~\ref{fig:structure}(c). Under high pressure, water molecules are forced closer together, resulting in an increased population of interstitial water molecules that shift toward the first coordination shell, leading to more acute angles between the oxygen atoms. These deviations from ideal tetrahedral geometry demonstrate the significant structural distortions induced in liquid water under high-pressure conditions.

\subsection{\label{sec:tetrahedral}Pressure Effects on Dipolar Correlations and the Kirkwood Factor}

\begin{figure}
\includegraphics[width=3.5in,height=4.88in]{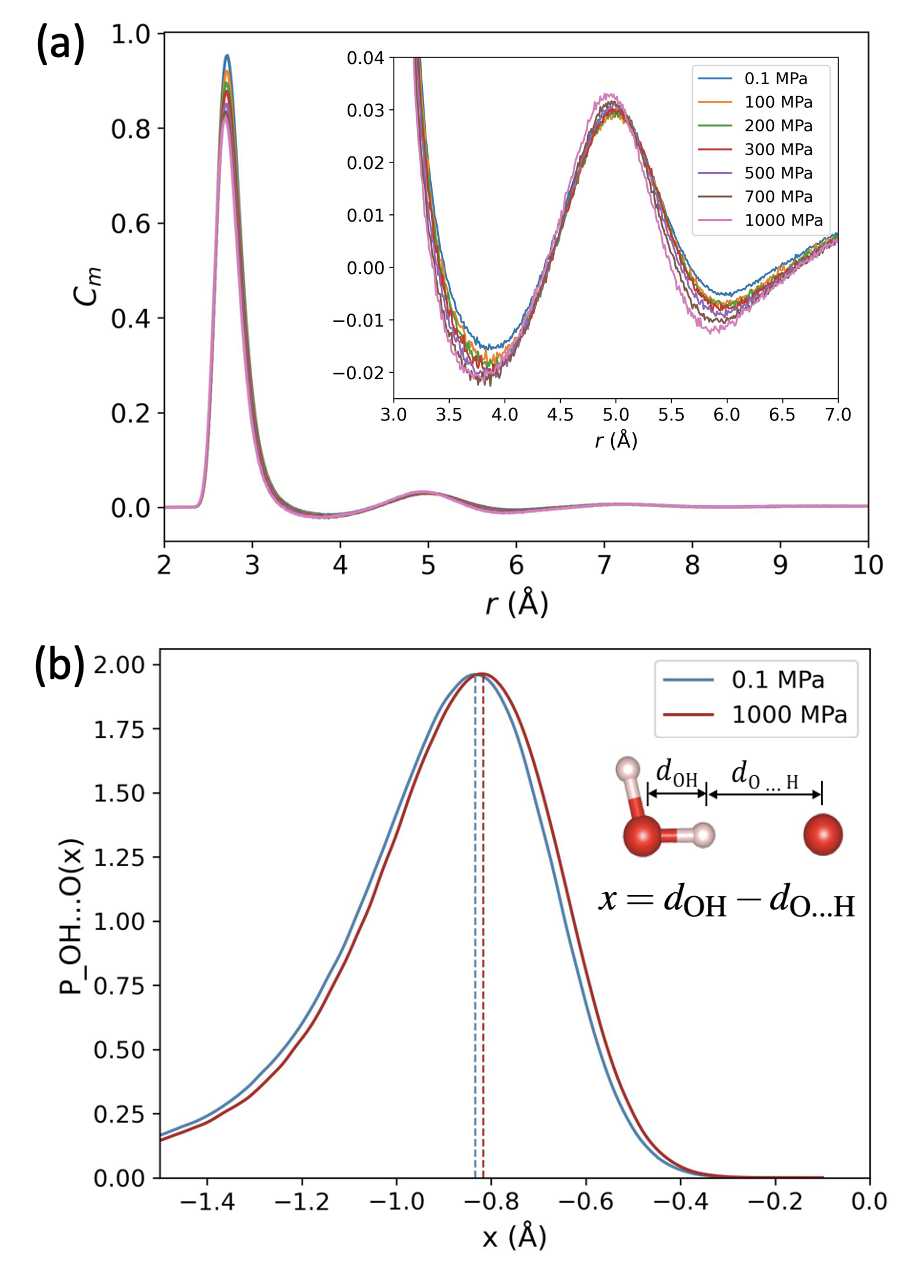}
\caption{\label{fig:analysis}
Analysis of water dipole correlation as a function of pressure at 330 K. (a) Dipolar correlation function \(c_m(r)\) across pressures, with an inset focusing on changes in the interstitious water and second peak between 3 Å and 7 Å. Several pressure conditions are omitted for visual clarity. (b) Spatial distribution of proton transfer coordinate \(x\) for water molecules under 0.1 MPa and 1000 MPa, where \(x = d_{\text{OH}} - d_{\text{O...H}}\).
}
\end{figure}

The structural changes induced by pressure in liquid water significantly affect dipolar correlations within the hydrogen-bond network. The dipolar correlation function \( C(\mathbf{r}) \) is defined as \( C(\boldsymbol{r}) = \langle\boldsymbol{d}(\mathbf{0}) \cdot \boldsymbol{d}(\boldsymbol{r})\rangle \), which describes the spatial correlation between dipolar density as a function of the distance \(\boldsymbol{r}\). Due to the discrete nature of water molecules, the dipolar density \( \boldsymbol{d}(\boldsymbol{r}) \) can be expressed as \(
\boldsymbol{d}(\boldsymbol{r}) = \sum_{i=1}^N \hat{\boldsymbol{\mu}}_i \delta\left(\boldsymbol{r}-\boldsymbol{r}_i\right),
\)
where \( \boldsymbol{r}_i \) and \( \hat{\boldsymbol{\mu}}_i \) is the position and the unit vector of the dipole moment of the \( i \)-th water molecule, respectively. The Kirkwood correlation factor \( G_{K} \), which quantifies the total angular correlations among the water dipoles, is obtained by integrating \( C(\boldsymbol{r}) \):

\begin{eqnarray}
G_{K} = \int C(\mathbf{r}) d\mathbf{r} = \frac{1}{N} \sum_{i=1}^N \sum_{j=1}^N \hat{\boldsymbol{\mu}}_i \cdot \hat{\boldsymbol{\mu}}_j.
\label{eq:GK}
\end{eqnarray}

We then define the dipole pair-correlation function \( c_m(r) \), which describes the correlation between dipole orientations, as \( \rho c_m(r) = C(r) - \left\langle \hat{\boldsymbol{\mu}}^2 \right\rangle \delta(\mathbf{r}) \), where \( \rho \) is the molecular number density. Therefore, \( G_{K} \) can be rewritten as:

\begin{eqnarray}
G_{K} &=& \int \left(\rho c_{m}(r) + \left\langle \hat{\boldsymbol{\mu}}^{2} \right\rangle \delta\left( \mathbf{r} \right)\right) d{r} \nonumber \\
&=& 1 + \rho \int c_{m}(r) d{r},
\label{eq:correlation}
\end{eqnarray}
indicating that \( G_{K} \) depends on both the system's density and the dipole pair-correlation function. Despite the increase in density with pressure, the reduction in \( G_{K} \) indicates a weakening of angular correlations among the water dipoles. 

As shown in Fig.~\ref{fig:analysis}(a), we present the simulated dipolar pair-correlation function \( c_m(r) \). Under ambient conditions, the directional nature of hydrogen bonding leads to the alignment of water dipoles in the first coordination shell with the central dipole, as these molecules are directly hydrogen-bonded to the central molecule. This alignment results in the first sharp peak observed in \( c_m(r) \). In the second coordination shell, although water molecules are not directly hydrogen-bonded to the central molecule, their dipole orientations are still influenced by the central dipole through the extended tetrahedral hydrogen-bond network. This interaction gives rise to the second peak in \( c_m(r) \) at approximately \( 5 \, \text{\AA} \). 
As the distance from the central molecule increases, thermal fluctuations and the inherent disorder in liquid water reduce the correlation between the dipoles of distant molecules and the central molecule, resulting in a gradual decay of the correlation peaks beyond the second coordination shell. Furthermore, interstitial water molecules, which occupy the space between the first and second coordination shells without forming direct hydrogen bonds with the central water molecule, tend to adopt slightly anti-parallel dipole orientations relative to the central dipole. This anti-parallel orientation generates a negative correlation in the interstitial region.

The effect of pressure on dipolar correlations closely relates to both the strength of directional hydrogen bonds and the tetrahedral order within the hydrogen-bonding network. As pressure increases, intermolecular distances decrease, leading to a more compact molecular arrangement and stronger hydrogen bonds. This effect is evidenced by the proton transfer coordinate shown in Fig.~\ref{fig:analysis}(b), where the centers of hydrogen bonds shift closer to their acceptors under pressure, indicating strengthened hydrogen bonding. This strengthening is consistent with the observed increase in dipole moment with pressure, which would typically enhance dipolar correlations. However, distortions in the tetrahedral structure of liquid water counteract this effect, leading to weaker angular correlations among water dipoles. 

At lower pressures, \( c_m(r) \) exhibits more pronounced peaks and valleys, indicating stronger dipolar correlations. However, as pressure increases, the first peak in \( c_m(r) \) diminishes significantly, reflecting a weakening of correlation, particularly between the central water molecule and those in its first coordination shell. This reduction in correlation contributes notably to the overall reduction in the Kirkwood correlation factor \( G_{K} \), and can be attributed to the increased population of interstitial water molecules between the first and second coordination shells, as shown in Fig.~\ref{fig:structure}(b). These interstitial molecules introduce additional interactions that disrupt the alignment of dipoles, causing the correlations between the central water molecule and its neighbors to weaken. Furthermore, the anti-correlations between interstitial and central water molecules become more pronounced under high-pressure conditions, further contributing to the overall decrease in the Kirkwood correlation factor \( G_{K} \). Thus, despite the strengthening of individual hydrogen bonds under pressure, the structural distortions leads to a less parallel alignment of dipoles, resulting in an overall weakening of dipolar correlations. 

\section{Conclusion}

In this study, we investigated the pressure-induced structural and dielectric changes in liquid water at room temperature over a pressure range of 0.1 MPa to 1000 MPa, using DNNs trained on data from DFT calculations. We observe a nonlinear increase in the static dielectric constant \(\varepsilon_{0}\) of water as pressure increases, which is qualitatively consistent with experimental observations. The primary factor driving this increase is the nonlinear rise in number density \( \rho \). As pressure increases, the liquid becomes more compressed, resulting in a higher number density of water molecules. This leads to stronger intermolecular interactions and larger fluctuations in the total dipole moment, thereby enhancing the dielectric response. Meanwhile, the shortening of O-O distances under pressure strengthens hydrogen bonds, contributing to a slight increase in the molecular dipole moment \( \mu \), which further increases \(\varepsilon_{0}\). Furthermore, the electronic contribution \(\varepsilon_{\infty}\), exhibits a modest increase, which can be attributed to the enhanced strength of interband transitions right across the bandgap, contributing minimally to the overall dielectric constant. 

Despite the increase in \(\varepsilon_{0}\), our results reveal a decrease in the Kirkwood correlation factor \(G_K\). This decrease is primarily due to pressure-induced distortions in the hydrogen-bonding network, which disrupt the tetrahedral order of water dipoles and weaken the angular correlations among them. These distortions are associated with the increased population of interstitial water molecules, which introduce additional interactions that disrupt the alignment of dipoles. Consequently, despite the strengthening of direct hydrogen bonds under pressure, the alignment of dipoles becomes less parallel due to these structural distortions, resulting in an overall weakening of dipolar correlations. Thus, the pressure-induced changes in the static dielectric constant \(\varepsilon_{0}\) are governed by a competition between the increasing density, which drives the increase in dielectric constant, and the diminishing dipolar correlations, which counteract this increase. 

Overall, this work provides crucial insights into the complex interplay between structural changes and dipolar correlations in liquid water under pressure, with broad implications for fields such as materials science, geophysics, and environmental engineering. Furthermore, the computational framework presented here offers a universal approach for studying the dielectric properties of other molecular fluids.

\begin{acknowledgments}
This research was supported by the Computational Chemical Center: Chemistry in Solution and at Interfaces, funded by the U.S. Department of Energy (DOE) under Award No. DE-SC0019394. Computational resources were provided by the National Energy Research Scientific Computing Center (NERSC), a DOE Office of Science User Facility, operated under Contract No. DE-AC02-05CH11231.
\end{acknowledgments}

\section*{Data Availability Statement}

The data that support the findings of this study are available from the corresponding author upon reasonable request.

\appendix

\nocite{*}
\bibliography{aipManuscript}% Produces the bibliography via BibTeX.

\end{document}